\documentclass[
    ,final            
  ]
  {aipproc}

\layoutstyle{6x9}

\newcommand{\sigmav}{<\sigma\,v>}
\newcommand{\cms}{cm^3s^{-1}}

\begin{document}

\title{Searches for Particle Dark Matter with gamma-rays}


\author{Jan Conrad}{
  address={Oskar Klein Centre, Physics Department, Stockholm University}
}

\begin{abstract}
In this contribution I review the present status and discuss some prospects for indirect detection of dark matter with gamma-rays. Thanks to the Fermi Large Area Telescope, searches in gamma-rays  have reached sensitivities that allow to probe the most interesting parameter space of the weakly interacting massive particles (WIMP) paradigm.  This gain in sensitivity is naturally accompanied by a number of detection claims or indications, the most recent being the claim of a line feature at a dark matter particle mass of $\sim$ 130 GeV at the Galactic Centre, a claim which requires confirmation from the Fermi-LAT collaboration and other experiments, for example HESS~II or the planned Gamma-400 satellite. Predictions for the next generation air Cherenkov telescope, Cherenkov Telescope Array (CTA), together with forecasts on future Fermi-LAT constraints arrive at the exciting possibility that the cosmological benchmark cross-section could be probed from masses of a few GeV to a few TeV. Consequently, non-detection would pose a challenge to the WIMP paradigm, but the reached sensitivities also imply that --optimistically-- a detection is in the cards.\\

\end{abstract}

\maketitle


\section{Introduction}
Observations from Galactic to cosmological scales leave little room to explanations for Dark Matter (DM) other than that it is provided by a new type of particle. The most studied option for this particle is a Weakly Interacting Massive Particle (WIMP). The reason for this is sometime referred to as the ``WIMP miracle'': for a particle of mass of the weak scale ($\sim$100 GeV), the annihilation cross section is given approximately by: $<\sigma v>  \sim \frac{\alpha^2}{M^²_{WIMP}} \sim  10^{-25} \cms $ and at the same time $\Omega_{relic} \sim  \frac{10^{-26} \cms}{\sigmav}$, from considering the relic density of a stable particle produced thermally in the early Universe\footnote{we will in this contribution concentrate on WIMP DM and leave other candidates, in particular axions or axion-like particles, aside.}.  WIMPs are searched for in a variety of ways: by particle production at accelerators, by searching for signals of nuclear recoils (direct detection) and last but not least by searching for a signal from secondary products of WIMP annihilation or decay, in particular gamma-rays.\\

\noindent  
Spectral signatures in gamma-rays can be classified in smoking gun signals and ambiguous signals. The smoking gun signals are mainly strong spectral features, such as bumps from virtual internal brems-strahlung \cite{Bringmann:2007nk} or lines from loop processes, e.g. \cite{Bergstrom:1988fp} \cite{Ullio:2002pj}, ambiguous signals are due to continuum emission from pion decay, e.g. \cite{Cesarini:2003nr} or hard power-laws (e.g. from final state radiation, e.g.  \cite{Birkedal:2005ep}).\\

\noindent
Often constraints in the annihilation cross-section, mass plane of WIMPs are presented under simplifying assumptions on the final state. Specifically,  annihilation into quarks is presented together with annihilation into leptons. Annihilation into quarks yields spectral shapes which are relatively independent of flavor (e.g. \cite{Fornengo:2004kj}), which is why constraints assuming --say -- 100 \% branching fraction into $b \bar{b}$ can be taken to be representative for a large class of models. A more accurate approach is provided by considering underlying models providing the WIMP, e.g. Supersymmetry, as is done in in a fit combining several observables in e.g. \cite{Scott:2009jn}\cite{Ripken:2010ja} or considering a large set of individual models, e.g. \cite{Cotta:2011pm}. Constraints presented here are for predominant annihilation into quarks.\\

\noindent
Gamma-rays are detected in space by pair conversion telescopes, such as the Fermi Large Area Telescope (Fermi-LAT)\cite{Atwood:2009ez}   and on ground by imaging air Cherenkov telescopes (IACT), such as HESS \cite{HESS}, VERITAS \cite{VERITAS} and MAGIC \cite{MAGIC},  and water Cherenkov telescopes, such as MILAGRO \cite{MILAGRO}. Most relevant for searches of Dark matter are the Fermi-LAT and IACTs, which are complementary in energy coverage. Fermi-LAT is mostly sensitive below $\mathcal{O}$(100 GeV) and is a survey instrument with large field of view ($\sim$ 2.4 sr), but with modest effective area ($\sim 1 $m$^2$). IACTs are pointing instruments are sensitive mostly to energies above  $\mathcal{O}$(100 GeV), they have a small field of view ($\sim$ 0.003 sr), but effective areas of $\sim 10^4 $m$^2$.

\section{Targets}
The number of photons created by DM particle annihilation is proportional to the square of the DM density along the line of sight.  N-body simulations show that our visible Galaxy is embedded in a much larger roughly spherical halo, which however shows a lot of substructure, e.g. \cite{Springel:2008cc}. 
The density is largest at the center of the halo, which is why the Galactic Centre should provide the largest signal from DM annihilation. Substructure reveals itself by close-by dwarf galaxies, but N-body simulations usually predict more abundant substructure than observed by ``barionic condensates'', i.e. the existence of dark matter-only  satellites is conjectured.   In addition, integrating over density enhancements may yield a Galactic diffuse contribution. The most interesting extragalactic source-type is Galaxy clusters, that might yield a signal due to strong substructure enhancement, e.g.  \cite{Pinzke:2011ek}. Diffuse extragalactic emission, originating from integration over all DM halos throughout the Universe is another viable channel. Table \ref{tab:targets} summarizes the targets.\\

\noindent
While consideration of DM structure provides guidance regarding possible  targets for  searches, the remaining ignorance about the density distribution constitutes the most important uncertainty in constraining the particle properties of DM. The impact of this ignorance depends on the target: for dwarf galaxies relatively robust density profile estimates can be obtained based on measured stellar motions, whose uncertainties affect constraints by factors of a few \cite{Abdo:2010ex} or somewhat less for the combined source analysis \cite{Ackermann:2011wa}. For the Galactic halo and galaxy clusters uncertainties can be several orders of magnitude, inferred mainly from N-body simulations with relatively weak or indirect experimental constraints. If the very inner part of the Galaxy is excluded from the analysis, these uncertainties are somewhat reduced (factors of a few, see e.g. \cite{Abramowski:2011hc}).\\

\noindent
The present status of the most relevant constraints is summarized in figure \ref{fig:status}. Note however that none of them (except the
(Fermi-LAT combined dwarf analysis, see below) includes the uncertainties in the dark matter density.

\begin{table}
\begin{tabular}{l||r||l}
\hline 
\tablehead{1}{l}{b}{Target/Signature}  & \tablehead{1}{r}{b}{Comments} &\tablehead{1}{r}{b}{References\tablenote{not complete}} \\
\hline
Galactic Centre  &Strongest expected signal, but hard background sources &\cite{Acero:2010}\cite{Cohen-Tanugi:2009}\cite{Murgia:2012} \cite{F_sym}\\& Poorly understood diffuse background.&\cite{Hooper:2011ti}\cite{Ripken:2007} \cite{GC_sim}  \\& Uncertain DM distribution.\\ \hline
Dwarf Galaxies   &Weak signal & \cite{Abdo:2010ex}\cite{Ackermann:2011wa}  \cite{Aharonian:2007km}\\
& small background & \cite{Aharonian:2009a}\cite{Acero:2010zzt} \cite{Albert:2007xg}  \\& relatively well determined DM  distribution &\cite{Aliu:2008ny}\cite{Aleksic:2011jx}\cite{veritas:2010pja} \\ & best for constraints  & \cite{Aliu:2012ga}\\ \hline 
Galaxy Clusters & Boost factor gives potential for detection & \cite{Ackermann:2010qj}\cite{Abramowski:2012au} \cite{Zimmer:2011vy}
\\& DM distribution not well understood & \cite{Pfrommer:2012mm} \cite{Ando:2012vu} \cite{Ackermann:2010rg} \\&bad for constraints, maybe good for detection  & \cite{Aleksic:2009ir} \\ \hline 
Galactic halo & Spatial and spectral signature &\cite{Abramowski:2011hc}  \cite{Ackermann:2012rg}\cite{Hutsi:2010ai} \\ &  backgrounds not well constrained &\cite{Cirelli:2009dv} \cite{Fornasa:2011yb} \\& (in Fermi-LAT energy range).\\& most promising venue for IACTs & \\ \hline
Extragalactic & Dependent on many unknowns &\cite{Abdo:2010dk} \cite{Abazajian:2010zb} \cite{Yuan:2011yb}  \\& (DM model, EBL \tablenote{Extragalactic Background light} absorption, cosmology) & \cite{Ackermann:2012uf} \cite{Fornasa:2011yb} \\& good as target for anisotropy studies & \\ \hline
Lines & Smoking gun & \cite{Pullen:2006sy}\cite{Abdo:2010nc}\cite{Ackermann:2012qk}  \\& weak signal & \cite{Eldik} \cite{Weniger:2012tx} \cite{Su:2012ft} \\& potential instrumental systematics & \cite{Hektor:2012kc} \cite{Tempel:2012ey} \\ \hline   
\hline
\end{tabular}
\caption{Usual targets for searches for particle dark matter - and some comments.}
\label{tab:targets}
\end{table}

\noindent

\subsection{Dwarf galaxies}
\begin{figure}
\includegraphics[height=.35\textheight]{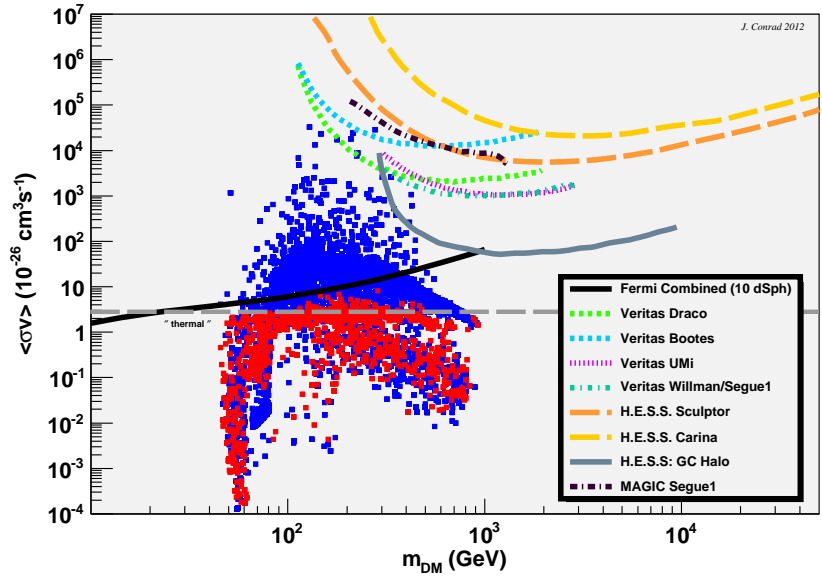}
\caption{Summary of current (most relevant) constraints on WIMP annihilation into gamma-rays. Points represent models of phenomenological Minimal Supersymmetry. Blue model points assume non-thermal dark matter production, red points are consistent with WMAP measurements. For details on these models see\cite {Abdo:2010ex}.}
\label{fig:status}
\end{figure}

\noindent
About 20 Dwarf spheroidal Galaxies, satellites of our own Galaxy have been targeted by the main gamma-ray experiments. Fermi-LAT presented observations of 14 Dwarf galaxies with DM constraints for about 2/3 of them based on availability of reasonable estimates of the DM distribution based on stellar data (see figure \ref{fig:status}) \cite{Abdo:2010ex}. H.E.S.S presented constraints from the Sagittarius Dwarf \cite{Aharonian:2007km} and Canis Major \cite{Aharonian:2009a} and Sculptor and Carina \cite{Acero:2010zzt}, MAGIC for Willman I and Draco \cite{Albert:2007xg} \cite{Aliu:2008ny} and Segue I \cite{Aleksic:2011jx}. Veritas for Bootes I, Draco, Ursa Minor and Willman I \cite{veritas:2010pja} and a -- using a deep ($\sim$ 50h) exposure for Segue I \cite{Aliu:2012ga}.  The typical exposure of the dwarfs for the ground-based experiments is at about 10-20 hours, which can be compared to an equivalent 1000 hours (11 month of survey mode) of Fermi, which therefore provides comparable constraints even at 1 TeV assumed particle mass. However,  it should be emphasized that in the region above 1 TeV the IACT will be more sensitive than the Fermi-LAT, thus truly complementary to Fermi-LAT.  Constraints derived from single dwarf analyses are about one order of magnitude larger than the interesting cosmological bench mark\footnote{see \cite{Steigman:2012nb} for a more accurate calculation of this quantity} $\sigma \, v \sim 3 \times 10^{26} cm^{3}s^{-1}$. It is also noteworthy, that it has recently been claimed that the H.E.S.S. constraint might be over-optimistic by almost one order of magnitude \cite{Viana:2011tq}, due to updated and more accurate estimates of its DM density.\\

\noindent
A particular interesting analysis approach is the stacking of dwarf galaxies, which due to the universality of the DM spectrum in dwarf galaxies, can be performed as a combined likelihood ("likelihood stacking") analysis \cite{Ackermann:2011wa}. Not only is the statistical power increased, but also the impact of individual uncertainties in dark matter density, is reduced. This approach resulted in the strictest limits on annihilation cross-section, yet arguably most robust, excluding the canonical cosmological cross-section up to masses of about 30 GeV. As mentioned previously, the analysis presented in \cite{Ackermann:2011wa} constitutes a step forward in one other aspect: the uncertainties in the DM density have been included in the derivation of the constraints in a statistically well established fashion (``profile likelihood'').

\subsection{Galactic Centre and halo}
The Galactic Centre (GC) is expected to be one of the strongest sources for gamma-ray radiation due to DM annihilation. At the same time, the GC is crowded with conventional gamma-ray sources: HESS and Fermi-LAT sources at the GC are consistent with each other and known sources \cite{Acero:2010}\cite{Cohen-Tanugi:2009}\cite{Murgia:2012}\cite{Hooper:2011ti}. 
H.E.S.S. has used the spectrum of the GC sources (which is inconsistent with being dominantly due to DM), to constrain DM models \cite{Ripken:2007}, which allows constraining models with extreme properties (for example those having a large contribution from virtual internal bremsstrahlung) and under assumptions of very cuspy profiles.\\

\noindent
The more promising approach is to exclude the very center from the analysis. However, for GeV energies the modeling of the diffuse emission in the inner galaxy poses a challenge, which is why Fermi-LAT has not presented Dark Matter constraints for the inner galaxy yet. For Galactic diffuse emission excluding the inner 15$^\circ$ $\times$ 15$^\circ$, an analysis attempting to quantitatively take into account the uncertainties in the cosmic ray induced Galactic diffuse emission obtaining competitive constraints has been presented in \cite{Ackermann:2012rg}.\\

\noindent
At IACT energies the systematics of the diffuse emission become less important. The analysis presented in 
\cite{Abramowski:2011hc} applies an on-off technique, where (as usual in IACTs) the background is determined by off-source observations. In this particular case, the off source region is defined within the field of view, at distances greater $\sim 0.1$ kpc from the Galatctic Center. The signal region excludes the very center, i.e. distances smaller than $\sim$ 0.05 kpc. This excludes not only the strong Galactic Centre sources, but also makes the constrain relatively insensitive (within a factor 2) to the assumptions on DM profile.  The background estimate (except for instrumental systematics) is sensitive to gradients in Galactic Diffuse emission predictions only, which should be negligible at these energies.  As the GC is observed for almost 100 hours, and the expected signal is expected to be comparably large, the obtained constraints obtained are up to one order of magnitude better than for dwarfs, and the observation of the GC halo seems to be the most promising venue for IACT for the moment.\\

\subsection{Dark Matter with the Cherenkov Telescope Array}
\begin{figure}
  \includegraphics[height=.35\textheight]{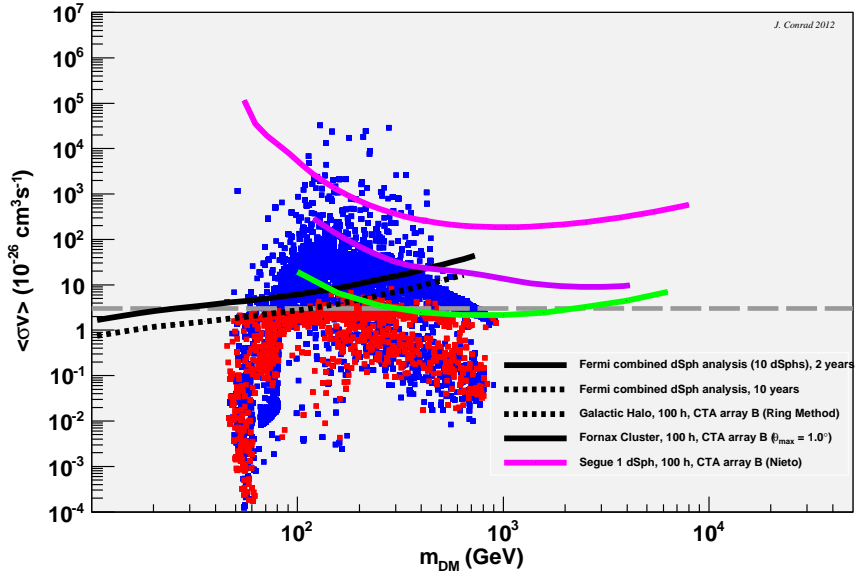}
  \caption{Same as in figure 1, except that predictions for CTA are presented together with a (pessimistic) expectation for the Fermi-LAT.}
\label{fig:CTA_forecast}
\end{figure}

The Cherenkov Telescope Array (CTA) will be the next generation Imaging Air Cherenkov Telescope \cite{Consortium:2010bc}. CTA will cover over 3 decades in energy, from a few tens of GeV up to several tens of TeV. The energy resolution and pointing resolution will be improved by a factor 2 to 3. The field of view will be up to 10 degrees (as compared to about 5 degrees for present day instruments). A comprehensive study of CTAs capabilities with respect to fundamental physics has been presented in \cite{Doro:2012xx}. As is the case for present day experiments, the most sensitive target will be Galactic Center halo. The forecast is that the thermal WIMP cross-section will be constrained from $\sim$ 200 GeV up to several TeV, see figure \ref{fig:CTA_forecast}.

\subsection{Dark Matter lines}
Dark matter lines have been suggested as a smoking gun signal for DM roughly 15 years ago \cite{Bergstrom:1988fp}. Constraints on line emission have been presented using EGRET \cite{Pullen:2006sy} data, Fermi-LAT \cite{Abdo:2010nc}\cite{Ackermann:2012qk} and H.E.S.S. data \cite{Ripken:2007}\cite{Eldik}. However, very recently, the claim of an indication of line emission in Fermi-LAT data \cite{Weniger:2012tx} \cite{Su:2012ft} has drawn considerable attention also at this conference \cite{Weniger} \cite{Finkbeiner}. Using an analysis technique similar to  \cite{Abdo:2010nc}, but doubling the amount of data as well as optimizing the region of interest for signal over square-root of background, \cite{Weniger:2012tx} found a (trial corrected) 3.2 $\sigma$ significant excess at a mass of  $\sim$ 130 GeV  that, if interpreted as a signal would amount to a cross-section of about $\sigmav \sim 10^{-27} \cms$. The signal is concentrated on the Galactic Centre with a spatial distribution consistent with an Einasto profile \cite{Bringmann:2012ez}. This is marginally compatible with the upper limit presented in \cite{Ackermann:2012qk}. The result was later confirmed by \cite{Su:2012ft}, who even claimed a (trial corrected)  5$\sigma$ detection significance.\\  

\noindent
The main challenge to this claim is the fact that the gamma-ray signal from the Earth's limb shows an excess of similar significance at the same energy, see e.g. Weniger (this conference). This effect appears after an appropriate zenith cut is applied that takes account for the fact that the Fermi-LAT is usually not head-on exposed to the Earth limb. Gamma-rays in the Earth limb are  caused by cosmic-ray interaction in the Earth atmosphere and their spectrum is featureless. The sample therefore constitutes a useful control sample for (not only) spectral feature searches. The community is now impatiently waiting for a verdict from the Fermi-LAT collaboration.  The main problem  at this point is the limited statistics in both the GC and limb sample. The Fermi-LAT has scheduled increase in statistics for the limb sample, by weekly 3 hour long Earth limb observations.\\

\noindent
Should an instrumental origin of the signal be ruled out (or results are inconclusive), Fermi-LAT could alter its observation mode to increase exposure to the Galactic Centre \cite{Su:2012ft}, on the other hand independent confirmation will be necessary. Both the operational HESS~II IACT and the future satellite mission Gamma-400 \cite{Galper:2011bc} seem to be sensitive enough to provide confirmation \cite{Bergstrom:2012vd}.

\subsection{Other searches for Particle Dark Matter}
\begin{figure}
 \includegraphics[height=.35\textheight]{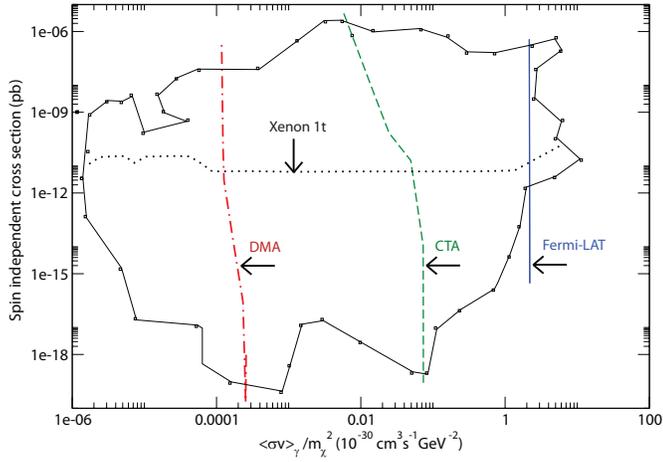}
  \caption{Contour representing the model-space (phenomenological Minimal Supersymmetry) occupied in direct detection cross section versus mass-squared scaled annihilation cross section. Lines represent expected sensitivities for the Xenon-1t direct detection experiment, as well as Fermi-LAT, CTA  and the speculative Dark Matter Array (DMA) (calculated in \cite{Bergstrom:2010gh}). Figure taken from \cite{Bergstrom:2012fi}.}
\label{fig:complementarity1}
\end{figure}

For indirect detection, constraining annihilation cross-section, gamma-rays are the most attractive probe, providing targets with high signal flux expectation and relatively well constrained astrophysical uncertainties. As this conference is about gamma-rays, the detailed discussion of results of direct searches and accelerator searches (as well as other indirect probes) is out of scope. However, it is worth pointing out the complementarity of the different probes once again. This complementarity is ultimately model-dependent, but at least for one of the most attractive candidates for a WIMP theory (Supersymmetry) it is apparent. In figure \ref{fig:complementarity1}, taken from \cite{Bergstrom:2012fi}, it is obvious that there is little correlation between scattering cross section and annihilation cross-section in phenomenological Minimal Supersymmetry: direct detection and neutrinos from the Sun\footnote{spin-dependent scattering cross section dominates the expected signal, as the capturing by the Sun is more important than annihilation.} constrain the model space approximately orthogonal to indirect detection by gamma-rays.\\

\noindent
Another example concerns collider searches for particles beyond the standard model. Collider detection and subsequent measurements of the sparticle mass spectrum and splitings of a supersymmetric model providing Dark matter might be used to calculate annihilation cross-sections and relic density \cite{Baltz:2006fm}. In the framework of phenomenological Minimal Supersymmetry, however, it is quite conceivable that only after a combination with indirect detection experiments the solution providing DM can be identified \cite{Bertone:2011pq}.

\section{Conclusions}
Since the launch of the Fermi-LAT indirect detection of Dark matter with gamma-rays has reached sensitivities that allow to probe the most interesting parameter space of the WIMP paradigm. The crucial target here are dwarf spheroidal galaxies that are analysed in a combined fashion. This gain in sensitivity is naturally accompanied by a number of detection claims or indications, the most recent being a line feature at a WIMP mass of $\sim$ 130 GeV at the Galactic Centre. In the near future, ut is now up to the Fermi-LAT Instrument team, as well as possible HESS (within a timescale of about 1-2 years) to confirm or refute this claim.\\

\noindent
IACTs have gained ground. The constraints obtained from observations of the Galactic Centre halo constitute a large step forward in terms of constraining WIMP parameter space above masses of about a TeV. Predictions for CTA, employing a Galactic Center halo analysis, together with forecasts on the Fermi-LAT constraints obtainable after 10 years arrive at the exciting possibility that the cosmological benchmark cross-section could be probed from masses of a few GeV to about 10 TeV.\\

\noindent
As the popularity of the WIMP as dark matter candidate rests in large part on the ``WIMP miracle'', it is fair to claim that a non detection in the coming 10 years will imply a serious challenge to the paradigm.  Or to put it more optimistically: a discovery within the next 10 years is certainly in the cards.



\begin{theacknowledgments}
The author thanks Lars Bergstr\"om and Joakim Edsj\"o for useful discussions. Christian Farnier is thanked for providing the model points in figures 1 and 2. JC is a Research 
Fellow of the Royal Swedish Academy of Sciences financed by a grant of the Knud and Alice Wallenberg foundation.
\end{theacknowledgments}



\bibliographystyle{aipproc}   


\IfFileExists{\jobname.bbl}{}
 {\typeout{}
  \typeout{******************************************}
  \typeout{** Please run "bibtex \jobname" to optain}
  \typeout{** the bibliography and then re-run LaTeX}
  \typeout{** twice to fix the references!}
  \typeout{******************************************}
  \typeout{}
 }


\end{document}